\documentclass[twocolumn,showpacs,preprintnumbers,amsmath,amssymb]{revtex4}

\usepackage{graphicx}
\usepackage{dcolumn}
\usepackage{bm}

\begin{document}

\title{Exactly Solvable Model for Helix-Coil-Sheet Transitions in Protein Systems}

\author{John S. Schreck}
\email{jss74@drexel.edu}
\author{Jian-Min Yuan}%
\email{Yuan@drexel.edu}

\affiliation{Department of Physics, Drexel University, Philadelphia, PA 19104}%

\date{\today}

\begin{abstract}
In view of the important role helix-sheet transitions play in protein aggregation, we introduce a simple model to study secondary structural transitions of helix-coil-sheet systems using a Potts model starting with an effective Hamiltonian. This energy function depends on four parameters that approximately describe entropic and enthalpic contributions to the stability of a polypeptide in helical and sheet conformations. The sheet structures involve long-range interactions between residues which are far in sequence, but are in contact in real space. Such contacts are included in the Hamiltonian. Using standard statistical mechanical techniques, the partition function is solved exactly using transfer matrices. Based on this model, we study thermodynamic properties of polypeptides, including phase transitions between helix, sheet, and coil structures. 
\end{abstract}

\maketitle
In late the 1950s and early 1960s, Zimm and Bragg (ZB) and Lifson-Roig (LR) studied helix-coil transitions of simple models of homopolypeptides by employing rigorous statistical methods based on partition functions and transfer matrices~\cite{zimm_theory_1959}. In the 1970s and 1980s, these models were extended to include copolymers and medium-ranged interactions, and were used to characterize the experimental results of all amino acids and many proteins~\cite{review_scheraga}.  Because of the close coupling between the theoretical and experimental studies, ZB, LR, and related models have stimulated much interest in helix-coil transitions~\cite{helix_book}, which is still an active field of research up to the present time~\cite{takano, germans_1}. For reviews, see Ref.~\cite{review_scheraga}.
  
However, conformation changes of polypeptides involving sheet structures, such as helix-sheet transitions, are not as well characterized as for helix-coil transitions. In the late 1970s, using a multi-state model, Tanaka and Scheraga~\cite{tanakasch} considered extended and chain-reversal states in addition to helix-coil transitions. In Ref.~\cite{wako}, medium-range interactions were taken into account to study helices, extended structures, and coils. More recently, Mattice and Scheraga~\cite{mattice_matrix_1984},  Sun and Doig~\cite{sun_and_doig}, Hong and Lei~\cite{hong_statistical_2008-1}, and others have included sheet structures in statistical models for homo-polypeptides. The difficulties in constructing models for sheets lie primarily in the interactions between residues that are long-range in sequence but are close in physical space, and in the rich variety of structures associated with sheets, turns, and loops, thus a large number of parameters required for their description. In this article, we introduce a simple statistical mechanical model for helix-coil-sheet transitions of homo-polypeptides, starting with an effective Hamiltonian. Instead of an Ising-like model, the treatment is built on a multi-state Potts model, which is capable of explicitly describing some of the long-range interactions exhibited by sheet structures. The objective is that this simple model extends the helix-coil treatments to protein systems with three or more secondary structures. 

An important step in a statistical mechanical approach like ZB, LR, Ising and Potts models is to construct the partition function for the system, based on which all thermodynamic properties are obtainable. As in ZB and LR models, partition functions factorize in terms of transfer matrices. However, ZB or LR theories start with a combinatorial partition function without defining an effective Hamiltonian. More generally, if an energy function $H(i)$ is defined, where $i=(i_{1}, \dots, i_{n})$ and $i_{n}$ is the micro-state of the \emph{nth} residue which could occupy one of $q$ possible states (conformations) labeled as $\{1, 2, \dots, q\}$, the partition function for a system of $N$ residues with periodic boundary conditions reduces to 
\begin{eqnarray}
\label{partition_function}
Z_{N} = \sum_{i_{1}=1}^{q} \sum_{i_{2}=1}^{q} \cdots \sum_{i_{N}=1}^{q} e^{-\beta H(i)} = Tr\left(T^{N}\right) 
\end{eqnarray}
where $\beta = (k_{B} T)^{-1}$, $k_{B}$ is Boltzmann's constant, and $Tr$ is the matrix trace operation. The dimension of a transfer matrix in a one-dimensional (1D) Ising model is $2\times 2$ and for a $q$-state Potts model, the dimension of a transfer matrix is $q\times q$. For Potts models with long-range interactions of range $L$ along a 1D chain, as Glumac and Uzelac~\cite{uzelac_1988} showed in their formulation, the dimension of a transfer matrix becomes $q^{L} \times q^{L}$. Eq.~(1) may be further simplified by diagonalizing the transfer matrix $T$. 
 
More recently, Hamiltonians of polypeptide chains have been described using a variety of Ising-like models ~\cite{takano, munoz1, bruscolini} and Potts models~\cite{anan3, goldstein}, and also using an \emph{ab initio} model~\cite{germans_1}. In particular, the WSME model~\cite{munoz1, bruscolini} uses two terms to construct an effective Hamiltonian and partition function: (1) the free energy term associated with the entropic cost of forming a pair of native residue conformations with restricted dihedral angles and (2) an enthalpic term associated with solvent-mediated contact energies between residues. Thus, residues may be either native or denatured, but not specific enough to distinguish sheets from helices. Our approach to polypeptides is based on a Potts model, where residues could assume many conformations including sheet, helix, coil, and turn. Before discussing the full helix-coil-sheet system, let us consider the simpler case of helix-coil transitions where an effective ($q=2$) Potts Hamiltonian (free energy in reality) can be written for a protein consisting of $N$ residues as 
\begin{equation}
\label{helix_coil_label}
-\beta H_{hc} =  h_{1} \sum_{n=1}^{N} \delta(1, i_{n})+ \beta J_{1} \sum_{n=k+1}^{N-1} \prod_{j=0}^{k-1} \delta(1,i_{n-j}) 
\end{equation}
where we assign $i_{n}=1$ to a residue in helix conformation and $i_{n}=2$ to a residue in coil conformation. The subscript `hc' in $-\beta H_{hc}$ means `helix-coil' and `1' in $h_{1}$ and $J_{1}$ refers to helix. The meanings of these parameters are similar to those described in the WSME model, where $h_{1}<0$ refers to an entropic cost from converting a coil to a helical residue, and $J_{1}>0$ refers to a contact energy between residues. In the present article, contact energies $J_{i}$ are free-energies associated with solvent-mediated interactions, including hydrogen bonds, van der Waals, polar interactions, etc. The Kronecker delta $\delta(1, i_{n})$ yields one if the $n$th residue is helical, and zero otherwise. 
In the second term of Eq.~(\ref{helix_coil_label}), the range $k$ determines the range of interaction. In $\alpha$-helices, where $k$ equals 3, residues at positions $n-3$, $n-2$, and $n-1$ are all helical when an H-bond forms between the $(n-4)th$ and $nth$ residues. Additionally, the $(n-4)th$ and $nth$ residues are not required to have the same conformation; in fact they could be in any conformation. When $k=1$, the effective Hamiltonian becomes $-\beta H_{hc}$ = $h_{1} \sum_{n=1}^{N} \delta(1, i_{n}) + \beta J_{1} \sum_{n=2}^{N} \delta(1, i_{n}) \delta(i_{n-1}, i_{n})$. The second term in Eq.~(\ref{helix_coil_label}) is also similar to the Hamiltonian of the GMPC model, which is a microscopic theory for helix-coil transitions based on a $q$-state Potts model~\cite{anan3, badasyan}. 

\begin{figure}
	\begin{center}
	\includegraphics[width=225pt]{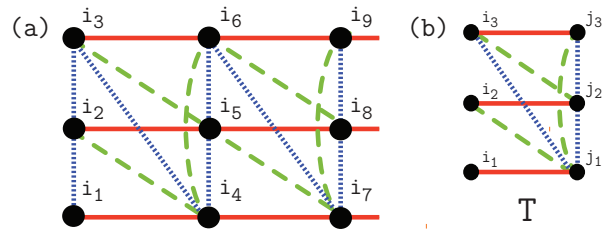}
	\caption{(Color online) a) Graphical representation of the partition function for the case L=3. The black dots mark the locations of particles along the chain. The dotted (blue) lines, $K_{1}$, are nearest neighbor interactions. The dashed (green) lines, $K_{2}$, are next nearest-neighbor interactions. Solid (red) lines, $K_{3}$, are the $L$=3 interactions. b) Graphical representation of the transfer matrix $T$.}
	\label{q=3pf}
	\end{center}
	\end{figure}

To write down an effective Hamiltonian suitable for $\beta$-sheets, we need to include in it interactions up to length $L$ along the polypeptide chain. Such a Hamiltonian can be constructed by adapting the long-range spin model of Glumac and Uzelac~\cite{uzelac_1988}. For a chain of $N$ spins, their Hamiltonian can be written as 
\begin{equation}
\label{hamil}
-\beta H =  \sum_{l=1}^{L} \sum_{n=1}^{N} \beta K_{l} \delta(i_{n},i_{n+l})
\end{equation}
where K$_{l}$ is distance-dependent. Fig.~\ref{q=3pf}(a) illustrates a graphical representation of the $L=3$ case and facilitates the construction of transfer matrices for long-range Potts systems. For Potts systems on a 1D lattice, Glumac and Uzelac grouped the spins along a chain into columns of height $L$, the longest interaction length, transforming a long-range problem of spin interactions into a short range one relating nearest-neighbor columns of height $L$~\cite{uzelac_1988, uzelac_1993}, illustrated in Fig.~\ref{q=3pf}(b). Each column of spins represents a vector that can take on one of $q^{L}$ possible states. The transfer matrix thus has dimension $q^{L} \times q^{L}$. The various lines in Fig.~\ref{q=3pf} represent interactions $K_{1}, K_{2}, \dots, K_{L}$ in Eq.~(\ref{hamil}), and contribute to the partition function when the arguments in the Kronecker delta's are equal. 

Two modifications are made to apply the Glumac-Uzelac method of constructing transfer matrices to a protein system. Fig. 2(a) illustrates a segment of an anti-parallel $\beta$-sheet, where interactions can occur between residues which are remote in relative chain position, but are nearby in space. This is what is meant by `long-range' in protein systems. Thus, the long-range nature of a protein system comes from labeling the residues according to the sequence order and does not come from the spatial distance between two residues. Even with the difference in the definition of long-range-ness, the Glumac-Uzelac method can be used in solving the protein problem. The strengths of interactions between each residue-residue pair are similar and not dependent on the relative chain position $l$. This is a main difference between our Hamiltonian (see Eq.~(5) below) and Eq.~(\ref{hamil}). For simplicity, in this article we shall consider all contacts between $\beta$-strands are of the same strength. In making this modification, Eq.~(\ref{hamil}) is recast as $-\beta H = \beta K \sum_{l=1}^{L} \sum_{n=1}^{N} \delta(i_{n},i_{n+l})$, which drops the $l$-dependence of $K$, but maintains the long-range nature of the Kronecker interactions. 
	\begin{figure*}
	\begin{center}
	\includegraphics[width = 450pt ]{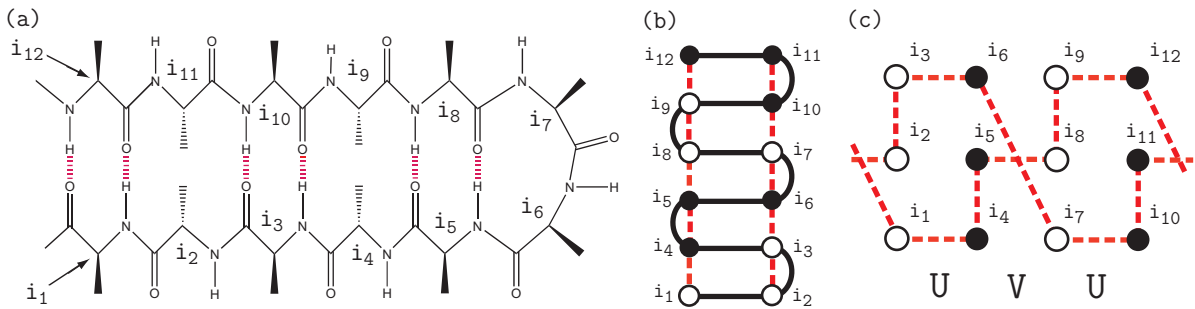}
	\caption{(Color online) a) A segment of an $L=11$ anti-parallel $\beta$-sheet chain. The sequence position of a residue is labeled and H-bonds are referenced by the dashed (red) lines. b) A simple pattern illustrating repeating $L=3$ and nearest-neighbor contact interactions, denoted by dashed (red) lines. The solid (black) lines represent peptide bonds. In (c), a diagram representing the partition function for the structure in (b). The first column in (c) are residues $i_{1}, i_{2}, i_{3}$, the second column are residues $i_{4}, i_{5}, i_{6}$, etc. Contacts are represented by dashed lines. The color of residues comprising the columns alternate in color from white to black, which corresponds to the residue pattern in (b). Repeated multiplication of matrices U and V generates the partition function for the whole chain.}
	\label{all_beta}
	\end{center}
	\end{figure*}

Secondly, according to Fig.~\ref{all_beta}(a), two hydrogen bonds form between residue-residue pairs, which occur for every other residue along a strand terminating at the turn. On the other hand, the residues along the $\beta$-strand that are not involved in hydrogen bonds with the opposite $\beta$-strand, could be involved in hydrophobic interactions with the opposite strand. To simplify the model, we assume that every residue-residue pair along neighboring strands forms contacts of the same strength, as stated above. The following pattern then represents H-bonding or hydrophobic interactions between two residues along neighboring strands, which we identify as contacts: $i_{1}\rightarrow i_{1+L}$, $i_{2}$$\rightarrow$$i_{1+L-1}$, $\cdots$ , $i_{(L+1)/2}$$\rightarrow$$i_{(L+1)/2+1}$. In the present work, the turn conformation is also counted as a sheet conformation, but, in principle, the model can be extended to include specifically turn conformations if $q>3$. The Kronecker delta's given in Eq.~(\ref{hamil}) are then modified to represent the aforementioned sheet-pattern. Additionally, for protein systems where the neighboring strands have the same interaction length $L$, the number of strands $M$, and the total number of residues $N$ are related by 
\begin{equation}
\label{nml}
N = M R, ~~~~R = \left(L+1\right)/2
\end{equation}
We can write the two-state effective Hamiltonian for a pattern such as the one in Fig.~\ref{all_beta}(a) extended for any $L$, while taking into account the two modifications made to Eq.~(3), as
\begin{eqnarray}
\label{coop_energy}
-\beta H_{sc} &=& h_{3} \sum_{n=1}^{N} \delta(3, i_{n})  \\ 
&+& \beta J_{3} \sum _{k=1}^{R} \sum _{m=1}^{M-1} b(i_{k,m}) \delta \left(i_{k+R (m-1)},i_{1-k+R (m+1)} \right), \nonumber\end{eqnarray}
where we denote $i_{n}=2$ (coil), or $3$ (sheet), $b(i_{k,m})$ $\equiv$ $\delta \left(3,i_{1-k+R(m+1)} \right)$ and only allows $J_{3}$ terms to accumulate when the residues at position $k+R(m-1)$ and $1-k+R(m+1)$ are locked in a sheet conformation and are in contact. The term $J_{3}>0$ now represents contacts between sheet residues, $h_{3}<0$ is the reduced entropic cost for coil to sheet conversions. The subscript `sc' in $-\beta H_{sc}$ refers to 'sheet-coil' and subscript `3'  in $h_{3}$ and $J_{3}$ refers to sheet. Unlike in Eq.~(\ref{helix_coil_label}), we do not require all residues between two residues in contact to be locked into the sheet state. 

To see the general pattern described by the second term in Eq.~(\ref{coop_energy}), we start by considering the simplest $L = 3$ case as shown in Fig.~\ref{all_beta}(b). In reality, the minimal structure in Fig.~\ref{all_beta}(b) may not even be considered as a sheet structure, but nevertheless illustrates the general behavior that the transfer matrix can be decomposed into a product of sub-transfer matrices. For $L=3$ case, the transfer matrix decomposes into a product of two matrices $U$ and $V$, as illustrated by Fig.~\ref{all_beta}(b) and (c). $U$ and $V$ are required to write out a general sequence of $M$ strands and are explicitly written with the help of Fig.~\ref{all_beta}(c) as
\begin{eqnarray}
\langle i \left| U \right| j \rangle &=& x^{\delta(i_{1},j_{1}) + \delta(i_{3}, j_{3})+\delta(j_{1}, j_{2})} \nonumber \\
\langle i \left| V \right| j \rangle &=& x^{\delta(i_{2}, j_{2})+\delta(i_{3}, j_{1}) + \delta(j_{2}, j_{3}) }
\end{eqnarray}
where $|i\rangle$ and $|j\rangle$ are neighboring column vectors of length $L$, where, for example, in Fig. 2(c), they can be $\left \langle i \right|$  = $\left \langle i_{1} i_{2} i_{3} \right|$ and $\left| j \right \rangle$ = $\left| i_{4} i_{5} i_{6} \right \rangle$, and $x = \exp\{\beta J_{3}\}$. Each transfer matrix $U$ and $V$ has dimension $q^{L} \times q^{L}$. This methodology works for any finite $L$, where the total number of transfer matrices needed to generate a periodic pattern for general $L$ is found to be equal to the total number of interactions over the distance $L+1$, which happens to equal the number $R$~\cite{schreck}. For example, for the $L=3$ case illustrated in Fig. 2(c), there are two interactions, a nearest-neighbor (for example, $i_{2}, i_{3}$, in Fig. 2(c)) and one over the longest range of interaction (for example, $i_{1}, i_{4}$, in Fig. 2(c)) thus two matrices are sufficient. 
       \begin{figure*}[t]
             \begin{center}   
             \includegraphics[width = 500pt]{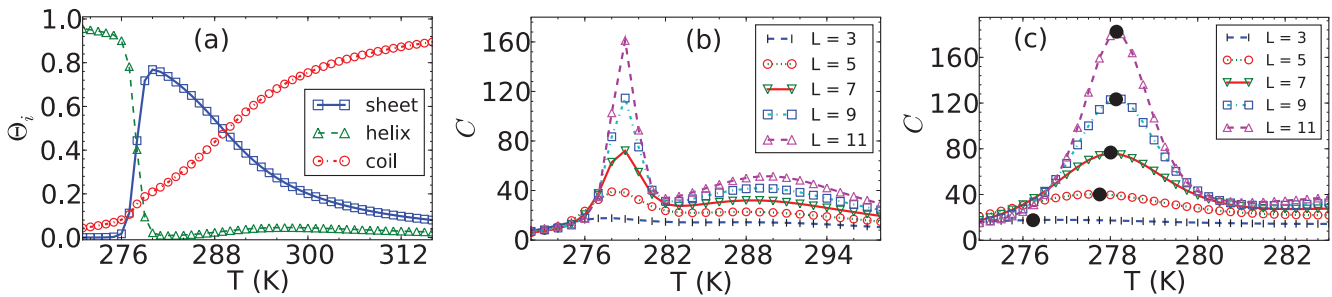}
             \caption{(Color online) All calculated quantities using $J_{1}$ = 2.85 kcal/mol, $J_{3}$ = 2.45 kcal/mol, $h_{1}$ = -4.91, and $h_{3}$ = -4.20. a) Order parameters for the case $L=11$ with $M=100$. b) Heat capacity (kcal/mol/K) vs. T for various strand lengths $L$ with $M=100$. c) The same plot as in (b) with more details of the helix-sheet transition given. Black dots denote transition temperatures which increase with range parameter $L$.}
             \label{all}
             \end{center}
             \end{figure*} 
For illustrating purposes, we explicitly consider a simple model of anti-parallel sheet-helix-coil systems, which starts with a three-state ($q=3$) effective Hamiltonian with four parameters that can describe transitions between sheet, helix, and coil structures. Helical conformations are assumed to form contacts between nearest neighbors only, that is, the $k=1$ case of Eq.~(2). The total effective Hamiltonian can be written as
\begin{equation}
-\beta H_{\text{hcs}} = -\beta H_{hc} - \beta H_{sc}
\end{equation}
where now $i_{n} = 1, 2$, or $3$, refers to helix, coil, and sheet, respectively, and the subscript `hcs' in $-\beta H_{\text{hcs}}$ refers to `helix-coil-sheet'. The partition function can be written in the form of Eq.~(\ref{partition_function}), when periodic boundary conditions are imposed, and calculated using transfer matrices, similar to the $L=3$ case as illustrated in Fig.~\ref{all_beta}(b) and (c). 
      
The parameters $h_{i}$ and $J_{i}$ are chosen so that the helix state is the most stable conformation at the lowest temperature in the interested temperature range. The coil dominates at high temperatures, where contact energies become relatively weak compared to thermal fluctuations. The sheet is thus an intermediary state~\cite{ding}. For some proteins, the sheet is seen as the most stable conformation at low temperature, where the helix conformation becomes an intermediary state~\cite{levy}. Our model can accommodate this case as well as a variety of others with proper choices of parameters. 
            
For systems with fixed numbers of residues, the partition function facilitates calculation of numerous thermodynamical quantities, such as the average energy, $\langle E \rangle$, the heat capacity, $C$, and the order parameters, $\Theta_{i}$, which are the average fractional content of $ith$ state among $q$ conformations at a particular temperature. To calculate the partition function, we choose a multi-stranded $\beta$-barrel system, which serves as an example of a protein system satisfying periodic boundary conditions. Inserting Eq.~(9) into Eq.~(1) and differentiating, we have for such a system
\begin{equation}
\small
C = \frac{\partial \langle E \rangle }{\partial T} = \frac{\partial}{\partial T}\left(k_{B} T^{2} \frac{\partial \ln Z_{N}}{\partial T} \right)~~\text{and}~~\Theta_{i} = \frac{\partial \ln\left(Z_{N} \right)}{\partial J_{i}},
\end{equation}
respectively. In Fig.~\ref{all}(a), the order parameters for helix, coil, and sheet are presented for the case $L=11$, $M=100$, and in Fig.~\ref{all}(b) and (c), we plot the temperature dependence of the heat capacity for various $L$ cases. The heat capacity curve show two peaks: the sharp, low-temperature peak signifies the helix-sheet transition, and the broad, high-temperature peak signifies the sheet-coil transition. These peak positions are approximately given by the crossing points of $\Theta_{i}$, shown in Fig. 3(a), between the helix and sheet and between the sheet and coil curves.  
                            
In conclusion, we have shown that, for a simple pattern associated with anti-parallel $\beta$-sheet structures, an effective Hamiltonian using a minimal number of parameters and its corresponding partition function can be constructed to study its helix-coil-sheet transitions. The partition function can be exactly computed by means of transfer matrices, which are used to calculate thermodynamical properties of the system, including the order parameters for helices and sheets and the heat capacity, which show that increasing strand length, $L$, plays a stabilizing role in the protein. 

We would like to acknowledge Zvonko Glumac and Katarina Uzelac for stimulating discussions and sending us unpublished results. JMY wants to thank Professor Sheng H Lin and Dr. A. N. Morozov for discussion and support at the early stage of this work. We thank the Pittsburgh Supercomputing Center for computing support.


\begin{thebibliography}{elsarticle-num}

\bibitem{zimm_theory_1959} B. H. Zimm and J. K. Bragg, J. Chem. Phys. \textbf{31}, 526 (1959); S. Lifson and A. Roig, J. Chem. Phys. \textbf{34}, 1963 (1961).
\bibitem{review_scheraga} H. A. Scheraga, J. A. Vila, and D. R. Ripoll, Biophysical Chemistry, \textbf{101-102}, 255 (2002). A. J. Doig, Biophysical Chemistry, \textbf{101-102}, 281 (2002). 
\bibitem{helix_book} D. Poland and H. A. Scheraga, Theory of Helix-Coil Transitions in Biopolymers (Academic Press, New York, 1970).
\bibitem{takano} M. Takano, K. Nagayama, and A. Suyama, J. Chem. Phys. \textbf{116}, 2219 (2002).
\bibitem{germans_1} A.V. Yakubovich, I.A. Solov'yov, A.V. Solov'yov, and W. Greiner, Eur. Phys. J. D. \textbf{46}, 227(2008).
\bibitem{tanakasch} S. Tanaka and H. Scheraga, Macromolecules \textbf{9}, 812 (1976); S. Tanaka and H. Scheraga, Macromolecules \textbf{10}, 9 (1977); S. Tanaka and H. Scheraga, Macromolecules \textbf{10}, 305 (1977). 
\bibitem{wako} H. Wako, N. Saito, and H. A. Scheraga, J. Protein Chem. \textbf{2}, 221 (1983).
\bibitem{mattice_matrix_1984} W. L. Mattice and H. A. Scheraga, Biopolymers \textbf{23}, 1701 (1984). 
\bibitem{sun_and_doig} J. K. Sun and A. J. Doig, J. Phys. Chem. B. \textbf{104}, 1826 (2000).
\bibitem{hong_statistical_2008-1} L. Hong and J. Lei, Phys. Rev. E. \textbf{78}, 051904 (2008); L. Hong, J. Chem. Phys. \textbf{129}(22), 225101 (2008).
\bibitem{uzelac_1988} Z. Glumac and K. Uzelac, J. Phys. A. \textbf{21}, L421 (1988); Z. Glumac and K. Uzelac, J. Phys. A. \textbf{22}, 4439 (1989).
\bibitem{munoz1} V. Mu\~noz, P. A. Thompson, J. Hofrichter, and W. A. Eaton, Nature (London) \textbf{390}, 196 (1997).
\bibitem{bruscolini} V. Mu\~noz, E. R. Henry, J. Hofrichter, and W. A. Eaton, Proc. Natl. Acad. Sci. U.S.A. \textbf{95}, 5872 (1998).; V. Mu\~noz and W. A. Eaton, Proc. Natl. Acad. Sci. U. S. A. \textbf{96}, 11311 (1999); P. Bruscolini and A. Pelizzola, Phys. Rev. Lett., \textbf{88}, 258101 (2002).
\bibitem{anan3} N. S. Ananikyan, Sh. A. Hajryan, E. Sh. Mamasakhlisov, and V. F. Morozov, Biopolymers, \textbf{30}, 357 (1990).
\bibitem{goldstein} R. E. Goldstein, Phys. Lett. \textbf{104A}, 285 (1984).
\bibitem{badasyan} A. V. Badasyan, A. Giacometti, Y. Sh. Mamasakhlisov, V. F. Morozov, and A. S. Benight, Phys. Rev. E., \textbf{81}, 021921 (2010). 
\bibitem{uzelac_1993} Z. Glumac and K. Uzelac, J. Phys. A. \textbf{26}, 5267 (1993).
\bibitem{schreck} J. Schreck and J.M. Yuan, \textit{To appear}
\bibitem{ding} F. Ding, J. M. Borreguero, S. V. Buldyrev, H. E. Stanley, and N. V. Dokholyan, Proteins \textbf{53}, 220 (2003).
\bibitem{levy} M. Andrec, A. K. Felts, E. Gallicchio, and R. M. Levy, Proc. Natl. Acad. Sci. USA \textbf{102}, 6801 (2005). 

\end{thebibliography}
\end{document}